\documentclass[showpacs,amssymb,aps,prd,floatfix]{revtex4}
\usepackage{amsmath}
\usepackage{amssymb}
\usepackage{graphicx}
\usepackage{accents}
\usepackage{pstricks,pst-node,pst-coil,pst-plot}
\usepackage{mathtools}
\usepackage{ulem}
\usepackage{mathtools}
\usepackage{ulem}
\usepackage{graphicx}
\usepackage{color}
\usepackage{graphicx}
\usepackage{color}
\usepackage{dcolumn}
\usepackage{graphicx}
\usepackage{color}
\usepackage{dcolumn}
\usepackage{bm}
\usepackage{mathrsfs}
\usepackage{graphicx}
\usepackage{color}
\usepackage{dcolumn}
\usepackage{bm}
\usepackage{slashed}
\usepackage{epstopdf}
\usepackage{slashed}
\usepackage{bm}
\usepackage{slashed}
\usepackage{dcolumn}
\usepackage{bm}
\usepackage{slashed}
\usepackage{color}
\usepackage{subfigure}

\newcommand{\be}{\begin{equation}}
\newcommand{\ee}{\end{equation}}
\def\mh{E_{\mbox{Hawking}}}
\def\eg{E_{\mbox{Hawking-Geroch}}}

\def\mhy{E_{\mbox{Hayward}}}
\def\Area{|\Sigma|}

\def\ringA{\accentset{\circ}{A}}
\newcommand{\beq}{\begin{equation}}
\newcommand{\eeq}{\end{equation}}
\newcommand{\bee}{\begin{equation*}}
\newcommand{\eee}{\end{equation*}}
\def\d{\mathrm{d}}

\begin{document}

\title{Hayward Quasilocal Energy of Tori}

\author{Xiaokai He}\email{sjyhexiaokai@hnfnu.edu.cn}
\affiliation{School of Mathematics and Computational Science, Hunan First Normal University, Changsha 410205, China}

\author{Naqing Xie}\email{nqxie@fudan.edu.cn}
\affiliation{School of Mathematical Sciences, Fudan University, Shanghai 200433, China}


\begin{abstract}
This paper is dedicated to the investigation of the positivity of the Hayward quasilocal energy of tori. Marginally trapped tori have nonnegative Hayward energy. We consider a scenario of a spherically symmetric constant density star matched to an exterior Schwarzschild solution. We show that any generic tori within the star, distorted or not, trapped or not, have strictly positive Hayward energy. Surprisingly we find analytic examples of `thin' tori with negative Hayward energy in the outer neighborhood of the Schwarzschild horizon. These tori are swept out by rotating the standard round circles in the static coordinates but they are distorted in the isotropic coordinates. Numerical results also indicate that there exist horizontally dragged tori with strictly negative Hayward energy in the region between the boundary of the star and the Schwarzschild horizon.
\end{abstract}

\maketitle

\section{Introduction}
Finding a suitable notion of quasilocal energy-momentum for finite spacetime domains at quasilocal level is one of the most challenging problems in classical general relativity \cite{Pe82}. Even though there was high expectation in the 1980's, this problem has proven to be surprisingly difficult and we have no ultimately satisfied expression yet \cite{Sz}. However, there are various `lists of criteria of reasonableness' in the literature among which expect that the quasilocal energy should be nonnegative under certain energy conditions \cite{ChY88}. This expectation is inspired by the successful proof of the positivity of the total gravitational energy \cite{ADM,S+YI,S+YII,Wi}.

The existing candidates for the quasilocal energy are mixture of advantages and difficulties \cite{Sz}.  In 1993, Brown and York introduced a notion of quasilocal energy following the Hamiltonian-Jacobi method \cite{BY}. The energy expression can be viewed as the total mean curvatures comparison in the physical space and the reference space. When the surface in question is a topological sphere with positive Gauss curvature and positive mean curvature, Shi and Tam proved that the Brown-York energy is nonnegative \cite{ST02}. One important feature of the Brown-York energy is that it requires a flat reference via isometric embedding. The issue of isometric embedding is a very beautiful and challenging problem in mathematics. However, it is difficult to compute the precise value of the energy of a generic surface for working relativists. Hawking had a different definition whose advantages are simplicity and calculability \cite{Ha}.

Let $(\widetilde{M},\tilde{g})$ be a spacetime. Assume that $(\Sigma,\sigma)$ is a spacelike closed 2-surface with the induced 2-metric $\sigma$. Consider the ingoing $(-)$ and outgoing $(+)$ null geodesic congruences from $\Sigma$. Let $\theta_{\pm}$ be the null expansions. Then the Hawking energy \cite{Ha} of the 2-surface $\Sigma$ is defined as
\begin{equation*}\label{Hawkingenergy}
\mh(\Sigma)=\frac{1}{8\pi}\sqrt{\frac{\Area}{16\pi}}\int_\Sigma (\mbox{Scal}_{\sigma} +\theta_{+}\theta_{-})\d\sigma.\end{equation*}
Here $\mbox{Scal}_{\sigma}$ is the scalar curvature and $|\Sigma|$ is the area of $\Sigma$ with respect to the 2-metric $\sigma$.

For a 2-surface $\Sigma$ embedded in a time-symmetric hypersurface $(M,g)$, the Hawking energy is equal to the Geroch energy \cite{Ge73}. In this case, we call it the Hawking-Geroch energy and the energy expression reads
\begin{equation*}
\label{HG}
\eg(\Sigma)=\frac{1}{8\pi}\sqrt{\frac{\Area}{16\pi}}\int_\Sigma (\mbox{Scal}_{\sigma} -\frac{1}{2}H^2)\d\sigma
\end{equation*}
where $H$ is the mean curvature of $\Sigma$ in the hypersurface $M$.

This quantity has a very nice monotonicity along the inverse mean curvature flow and it plays a key role in the proof of the Riemannian Penrose inequality \cite{HI01}. Unfortunately, the value of the Hawking-Geroch energy is too small in some sense. Even in the flat $\mathbb{R}^3$, $\eg(\Sigma)$ is strictly negative unless $\Sigma$ is round. This drawback is then corrected by Hayward \cite{SH1}. The idea was adding additional terms based on the double-null foliation and the energy becomes zero for any generic 2-surface in the flat Minkowski spacetime $\mathbb{R}^{3,1}$. Moreover, there exists a family of ellipsoids in the Schwarzschild spacetime whose Hawking-Geroch energy tends to $-\infty$ \cite{FK11} but whose Hayward energy converges to a finite positive value at spatial infinity \cite{HWX20}.

Recall that $\Sigma$ is a closed 2-surface in spacetime with the induced 2-metric $\sigma$. Consider the ingoing $(-)$ and outgoing $(+)$ null geodesic congruences from $\Sigma$. Let $\theta_{\pm}$ and $\sigma_{ij}^{\pm}$ be the expansions and shear tensors of these congruences respectively, and $\omega^k$ be the projection onto $\Sigma$ of the commutators of the null normal vectors to $\Sigma$. The Hayward quasilocal energy \cite{SH1} is defined as
\begin{equation*}
\mhy(\Sigma)=\frac{1}{8\pi}\sqrt{\frac{|\Sigma|}{16\pi}}\int_\Sigma \Big(\mbox{Scal}_{\sigma} +\theta_{+}\theta_{-} -\frac{1}{2}\sigma_{ij}^{+}\sigma_{-}^{ij}-2\omega^k \omega_k\Big)\d\sigma.\end{equation*}

The anoholonomicity $\omega^k$ is a boost-gauge-dependent quantity \cite{Sz}. From now on, we further assume that a 2-surface $\Sigma$ lies in the time slice in a static spacetime which yields the anoholonomicity $\omega^k$ vanishes. Then the Hayward energy can be further rewritten as
\be
\begin{split}\label{HH1}
\mhy(\Sigma)&=\sqrt{\frac{|\Sigma|}{16\pi}}\Big(\frac{1}{8\pi}\int_\Sigma\mbox{Scal}_{\sigma} \d\sigma - \frac{1}{16\pi}\int_{\Sigma}(H^2-2|\ringA|^2)\d\sigma\Big)\\
&=\sqrt{\frac{|\Sigma|}{16\pi}}\Big(\frac{1}{8\pi}\int_\Sigma\mbox{Scal}_{\sigma} \d\sigma - \frac{1}{8\pi}\int_{\Sigma}(H^2-|A|^2 )\d\sigma\Big)\\
&=\eg(\Sigma)+\frac{1}{8\pi}\sqrt{\frac{|\Sigma|}{16\pi}}\int_{\Sigma}|\ringA|^2 \d \sigma
\end{split}
\ee
where $\ringA_{ij}=A_{ij}-\frac{H}{2}\sigma_{ij}$ is the traceless second fundamental form of $\Sigma$ in $M$. It is obvious that the value of the Hayward energy of $\Sigma$ is greater than its Hawking-Geroch energy.

Most known results of the quasilocal energy in the literature are concerned with 2-surfaces with spherical topology. There seems to be no `\textit{a priori}' restriction on the topology of the surface. There do exist trapped surfaces or minimal surfaces with toroidal topology \cite{Hu96,KMMOX17,MX17}. The ideal construction of quasilocal energy should work for any closed orientable 2-surfaces \cite{Sz}.

Suppose that $\Sigma$ is a marginally trapped torus (MTT), i.e., the mean curvature vanishes $H=0$ and the genus is equal to $1$. By the Gauss-Bonnet theorem, one has
\bee
\mhy(\Sigma)=\frac{1}{8\pi}\sqrt{\frac{|\Sigma|}{16\pi}}\int_{\Sigma}|\ringA|^2 \d \sigma \geq 0.\eee

This paper is dedicated to the investigation of the positivity of the Hayward quasilocal energy of tori. In Section \ref{II}, we consider a scenario of a spherically symmetric constant density star matched to an exterior Schwarzschild solution \cite{O86}. In Section \ref{III}, we show that any generic tori within the star, distorted or not, trapped or not, have strictly positive Hayward energy. Section \ref{IV} gives the energy expression for tori in the exterior Schwarzschild region and proves that `thin tori'  must have positive Hayward energy. Section \ref{V} includes counterexamples. Surprisingly we find analytic examples of `thin' tori with negative Hayward energy in the outer neighborhood of the Schwarzschild horizon. These tori are swept out by rotating the standard round circles in the static coordinates but they are distorted in the isotropic coordinates. Numerical results also indicate that there exist horizontally dragged tori with strictly negative Hayward energy in the region between the boundary of the star and the Schwarzschild horizon. Conclusions are summarized in the last section.

\section{Basic Setting}\label{II}
We consider a spherically symmetric initial data $(M,g)$ for a constant density star matched to an exterior Schwarzschild solution, which presents a time slice in a static spacetime $(\widetilde{M},\tilde{g})$. In the isotropic coordinates $\{x^i\}$, the 3-metric $g$ reads
\be
\label{metricss}
\begin{split}
g&=\Phi^4 (R)  \big((\d x^1)^2+ (\d x^2)^2 +(\d x^3)^2\big)\\
&=\Phi^4 (R)  \big(\d R^2+R^2\d\Theta^2 +R^2\sin^2\Theta \d\varphi^2\big),
\end{split}
\ee
where
\be\label{PhiR}
\Phi (R)= \left\{
\begin{array}{rl}
\frac{(1+\beta R_0^2)^{3/2}}{\sqrt{1+\beta R^2}} \ \ \mbox{for $R\leq R_0$},\\
1+ \beta\frac{R_0^3}{R}\ \ \mbox{for $R >  R_0$}.
\end{array}
\right.
\ee
Here $R=\sqrt{(x^1)^2+(x^2)^2+(x^3)^2}$ is the spherical radial coordinate and $R_0$ is the Euclidean radius of the star.

The nonzero components of the Riemann curvature tensors of $g$ are
\be
\begin{split}\label{Riemanniso}
R_{1212}=R_{R\Theta R\Theta}&=-2R\Phi^2\big(-R(\Phi^\prime)^2+\Phi(\Phi^\prime+R\Phi^{\prime\prime}) \big)\\
R_{1313}=R_{R\varphi R\varphi}&=-2R\Phi^2\sin^2\Theta \big(-R(\Phi^{\prime})^2+\Phi(\Phi^\prime+R\Phi^{\prime\prime}) \big)\\
R_{2323}=R_{\Theta\varphi\Theta\varphi}&=-4R^3\sin^2\Theta \ \Phi^2\Phi^\prime \big(\Phi+R\Phi^\prime \big),
\end{split}\ee
where the prime $\prime$ denotes the differentiation with respect to $R$.

\begin{figure}[htp]
\begin{center}
\includegraphics[width=0.35\textwidth]{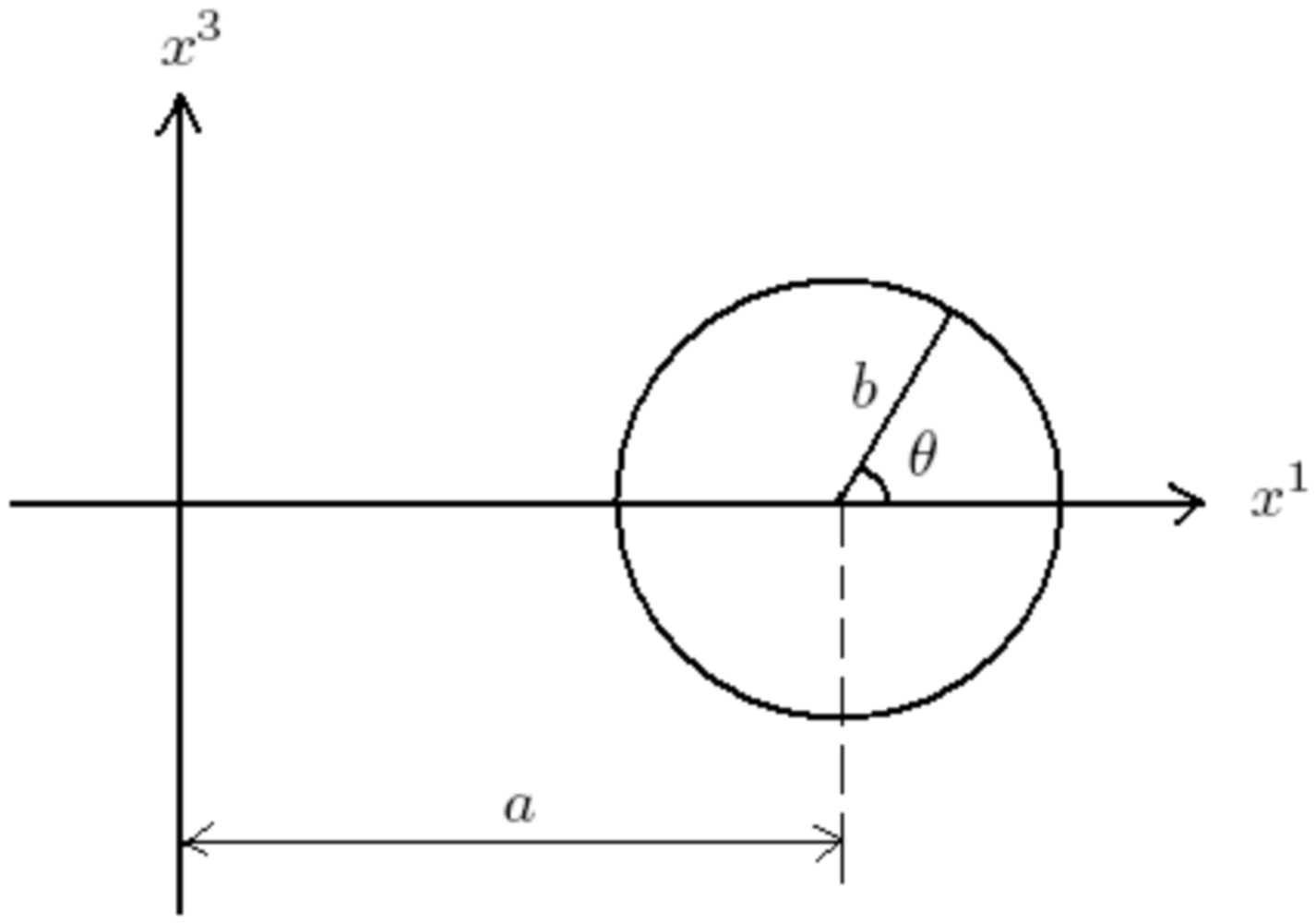}
\caption{\label{torus} \mbox{Standard circle in the $x^1-x^3$-plane.} }
\end{center}
\end{figure}

Assume that we are given a family of coordinate tori, denoted by $\Sigma$,  of major radius $a$ and minor radius $b$, with the $x^3$-axis as the symmetry axis. They are swept out by rotating the standard circles $\{(x^1,x^3)|(x^1-a)^2+(x^3)^2=b^2\}$ along the $x^3$-axis, as illustrated in Fig. \ref{torus}.

Indeed, these tori are parametrized as
\be\label{standardtoriiso}
\begin{split}
x^1&=(a+b\cos\theta)\cos\varphi,\\
x^2&=(a+b\cos\theta)\sin\varphi, \ \theta \in [0,2\pi), \ \varphi\in [0,2\pi)\\
x^3&=b\sin\theta.
\end{split}\ee
The induced 2-metric $\sigma$ reads
\bee\begin{split}
\sigma &=\Phi^4(R)\big(b^2\d \theta^2+(a+b\cos\theta)^2 \d \varphi^2\big)\\
&=\sigma_{22}\d \theta^2+\sigma_{33}\d \varphi^2
\end{split}\eee
where $\sigma_{22}=\Phi^4 b^2$ and $\sigma_{33}=\Phi^4 (a+b\cos\theta)^2$.
The area form of $\Sigma$ is
\bee\label{areaformiso}
\d \sigma=\sqrt{\det \sigma}\d\theta \wedge \d \varphi= \Phi^4 b(a+b\cos\theta)\d\theta \wedge \d \varphi.\eee
Denote by $\{e_2,e_3\}$ the orthonormal frame tangential to the surface $\Sigma$.
By the chain rule, one has
\bee
\begin{split}
e_2&=\frac{1}{\sqrt{\sigma_{22}}}\frac{\partial}{\partial
\theta}=\frac{1}{\Phi^2b}\bigg{(}
-\frac{ab\sin\theta}{\sqrt{a^2+b^2+2ab\cos\theta}}\frac{\partial}{\partial R}
-\frac{b(b+a\cos\theta)}{a^2+b^2+2ab\cos\theta}\frac{\partial}{\partial\Theta}\bigg{)}\\
&=(e_2)^R\frac{\partial}{\partial R}+(e_2)^\Theta\frac{\partial}{\partial\Theta},\\
e_3&=\frac{1}{\sqrt{\sigma_{33}}}\frac{\partial}{\partial
\varphi}=\frac{1}{\Phi^2 (a+b\cos\theta)}\frac{\partial}{\partial\varphi}\\
&=(e_3)^\varphi\frac{\partial}{\partial\varphi}.
\end{split}
\eee
Here
\bee\begin{split}
(e_2)^R &= -\frac{1}{\Phi^2}\frac{a\sin\theta}{\sqrt{a^2+b^2+2ab\cos\theta}} ,\\
(e_2)^\Theta & =-\frac{1}{\Phi^2}\frac{b+a\cos\theta}{
a^2+b^2+2ab\cos\theta},\\
(e_3)^\varphi&=\frac{1}{\Phi^2 (a+b\cos\theta)}.
 \end{split}\eee

\section{Tori within the Star}\label{III}
In this section, we prove that any generic topological tori lying entirely within the star must have positive Hayward energy. Without loss of generality, from now on we assume that $R_0=1$ for convenience. In this case, the conformal factor $\Phi=\frac{(1+\beta)^{\frac{3}{2}}}{\sqrt{1+\beta R^2}}$ for $R \leq 1$ and $\Phi=1+\frac{\beta}{R}$ for $R >1$. With this choice, the asymptotic mass of the spacetime is $m=2\beta$.

The contracted version of the Gauss equation for $(\Sigma,\sigma)$ in $(M,g)$ reads
\bee
\mbox{Scal}_{\sigma}-H^2+|A|_\sigma^2=\sum_{i,j=2,3}R(e_i,e_j,e_i,e_j).\eee
Here $\{e_2,e_3\}$ are the orthonormal frame of the tangent bundle of $(\Sigma,\sigma)$, $H$ and $A$ are the mean curvature and the second fundamental form of $(\Sigma, \sigma)$ in $(M,g)$ respectively.

By \eqref{HH1}, the Hayward quasilocal energy can be rewritten as
\bee\label{HHG}
\mhy(\Sigma)=\frac{1}{8\pi}\sqrt{\frac{|\Sigma|}{16\pi}}\int_{\Sigma}2 R(e_2,e_3,e_2,e_3) \d \sigma.
\eee
For a standard torus $\Sigma$ \eqref{standardtoriiso} lying entirely within the star, the conformal factor $\Phi=\frac{(1+\beta)^{\frac{3}{2}}}{\sqrt{1+\beta R^2}}$, and straightforward calculation shows that $R(e_2,e_3,e_2,e_3)=\frac{4\beta}{(1+\beta)^6}$. It immediately follows that the Hayward energy is positive, no matter the torus is trapped or not. Note that the mass density of the star is $\rho=\frac{3\beta}{2\pi(1+\beta)^6}$ \cite{O86} and the Hayward energy is proportional to the mass density. This is similar to the case of the small sphere in nonvacuum \cite{Ber}.

The positivity is indeed a much stronger result which can be extended for any generic tori, distorted or not. From \eqref{metricss}, \eqref{PhiR} and \eqref{Riemanniso}, it is easy to verify that
\bee
R_{ijkl}=\frac{4\beta}{(1+\beta)^6}\big(g_{ik}g_{jl}-g_{jk}g_{il}\big)=\frac{\mbox{Scal}_{\sigma}}{6}\big(g_{ik}g_{jl}-g_{jk}g_{il}\big).\eee
This indicates that the star has positive constant sectional curvature. The integrand $R(e_2,e_3,e_2,e_3)$ turns out to be the sectional curvature of the 3-metric $g$ with respect to the tangent plane of a generic torus.

\section{Tori in the Exterior Schwarzschild Region}\label{IV}
In this section, we calculate the Hayward energy for the coordinate tori \eqref{standardtoriiso} in the exterior Schwarzschild region. In this case, the conformal factor $\Phi=1+\frac{\beta}{R}$ and
\bee
R(e_2,e_3,e_2,e_3)= \frac{R\beta(a^2+4b^2+8ab\cos\theta+3a^2\cos 2\theta)}{(R+\beta )^6}\eee
where $R=\sqrt{(x^1)^2+(x^2)^2+(x^3)^2}$.

The Hayward energy for $\Sigma$ is
\be\label{massexiso}
\begin{split}
\mhy(\Sigma)&=\frac{1}{8\pi}\sqrt{\frac{|\Sigma|}{16\pi}}\int_{\Sigma} 2 R(e_2,e_3,e_2,e_3)\d\sigma\\
&=\frac{1}{4}\sqrt{\frac{|\Sigma|}{16\pi}}\int_0^{2\pi}
\frac{ 2b \beta  (a+b \cos \theta) \left(a^2+4 b^2+3 a^2 \cos 2 \theta+8 a b \cos \theta\right)}{\left(a^2+2 a b \cos \theta+b^2\right)^{3/2} \left(\sqrt{a^2+2 a b \cos \theta+b^2}+\beta \right)^2}\d\theta.
\end{split}
\ee
For small $b$, one has
\bee
\begin{split}
\mhy(\Sigma)&=\frac{1}{4}\sqrt{\frac{|\Sigma|}{16\pi}}\int_0^{2\pi}
\bigg{[}\frac{2\beta(1+3\cos2\theta)}{(a+\beta)^2}b
+\mathcal{O}(b^2)\bigg{]}d\theta\\
&=\frac{1}{4}\sqrt{\frac{|\Sigma|}{16\pi}}\bigg{[}
\frac{4\pi\beta b}{(a+\beta)^2}+\mathcal{O}(b^2)\bigg{]}.
\end{split}
\eee
This shows that `thin tori' in the exterior Schwarzschild region have positive Hayward energy.

\begin{figure}[htp]
\begin{center}
\includegraphics[width=0.60\textwidth]{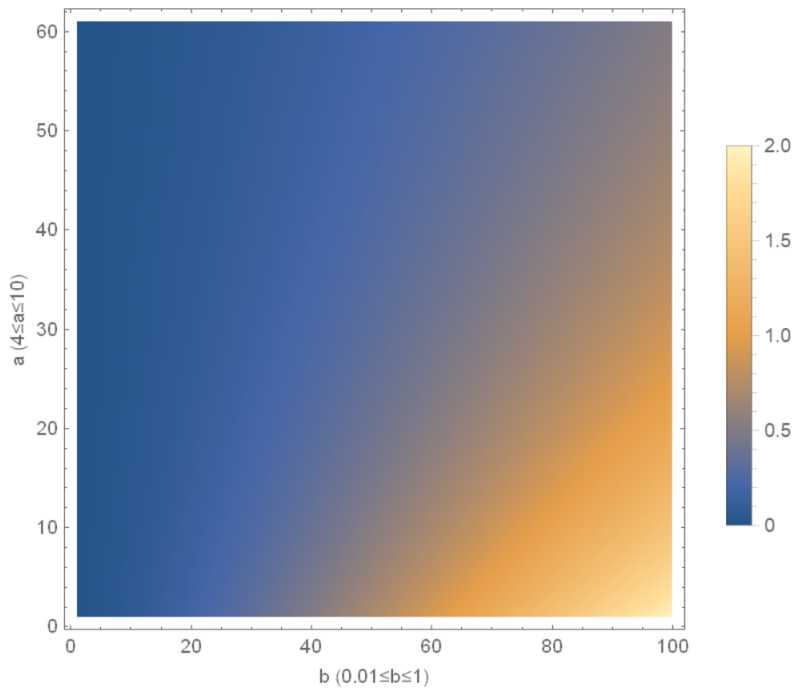}
\caption{\label{iso1} Plot of $\mhy(\Sigma)$ of the coordinate tori with respect to $a$ and $b$ for $\beta=3+2\sqrt{2}.$ }
\end{center}
\end{figure}

For general $a$ and $b$, it is not easy to figure out the triangle integral. We calculate the Hayward energy \eqref{massexiso} numerically and the results indicate that the Hayward energy is positive for the coordinate tori \eqref{standardtoriiso}.  For instance, we show the plot of $\mhy(\Sigma)$ with respect to $a\ (4\leq a\leq 10)$ and $b \ (0.01\leq b\leq 1)$ for $\beta=3+2\sqrt{2}$ in Fig. \ref{iso1}. The number $3+2\sqrt{2}$ is the minimal value of $\beta$ to construct marginally trapped tori in the star \cite[Section III]{KMMOX17}.

\section{Distorted Thin Tori with Negative Energy}\label{V}
In the previous sections, we have checked many cases that tori have positive Hayward energy. Does we really have confidence to conjecture that the positivity property for the Hayward quasilocal energy holds for tori in our scenario? Surprisingly, we find examples of distorted tori in the outer neighborhood of the Schwarzschild horizon with strictly negative Hayward energy. This could even be done in an analytic approach in the static coordinates.

The spatial Schwarzschild metric $g^m$ can be written in the static spherical coordinates $\{\hat{R}, \Theta, \varphi \}$ as
\bee g^m=\frac{1}{1-\frac{2m}{\hat{R}}}\d \hat{R}^2+\hat{R}^2\d\Theta^2 +\hat{R}^2\sin^2\Theta \d\varphi^2.
\eee
Denote by
\be\label{sphon}
\check{e}_1=\sqrt{1-\frac{2m}{\hat{R}}}\frac{\partial}{\partial\hat{R}}, \ \check{e}_2=\frac{1}{\hat{R}}\frac{\partial}{\partial \Theta}, \ \check{e}_3=\frac{1}{\hat{R}\sin\Theta}\frac{\partial}{\partial \varphi}\ee
the orthonormal frame. In this frame, the nonzero components of the Riemann curvature tensors are
\bee\label{curvature}
\begin{split}
R(\check{e}_1,\check{e}_2,\check{e}_1,\check{e}_2)&=R(\check{e}_1,\check{e}_3,\check{e}_1,\check{e}_3)=-\frac{m}{\hat{R}^3}\\
R(\check{e}_2,\check{e}_3,\check{e}_2,\check{e}_3)&=\frac{2m}{\hat{R}^3}.
\end{split}\eee
We consider a family of coordinate tori $\hat{\Sigma}$ which are parameterized as
\be
\begin{split}
y^1&=(a+b\cos\theta)\cos\varphi,\\
y^2&=(a+b\cos\theta)\sin\varphi, \ \theta \in [0,2\pi), \ \varphi\in [0,2\pi)\\
y^3&=b\sin\theta.\label{toricoordinate}
\end{split}\ee
Here $\hat{R}=\sqrt{(y^1)^2+(y^2)^2+(y^3)^2}$ and the tori are swept out by the standard circles $\{(y^1,y^3)|(y^1-a)^2+(y^3)^2=b^2\}$ along the $y^3$-axis.
It can be obtained from \eqref{toricoordinate} that the induced 2-metric $\hat{\sigma}$ on $\hat{\Sigma}$ is
\bee
\hat{\sigma} =\hat{ \sigma}_{22}\d \theta ^2+ \hat{\sigma}_{33}\d \varphi^2\eee
where
\bee
\begin{split}
\hat{\sigma}_{22}&=\frac{b^2\big{(}(
a^2+b^2+2ab\cos\theta)^{\frac{3}{2}}
-2m(a\cos\theta+b)^2\big{)}}{
(a^2+b^2+2ab\cos\theta)^{\frac{3}{2}}-
2m(a^2+b^2+2ab\cos\theta)},\\
\hat{\sigma}_{33}&=(a+b\cos\theta)^2.
\end{split}\eee
The area form of $\hat{\Sigma}$ with respect to the induced 2-metric $\hat{\sigma}$ is
\bee
\begin{split}
\d\hat{\sigma}&=\sqrt{\det \hat{\sigma}}\d\theta\wedge \d\varphi\\
&=b(a+b\cos\theta)\sqrt{\frac{(a^2+b^2+2
ab\cos\theta)^{\frac{3}{2}}
-2m(a\cos\theta+b)^2}{
(a^2+b^2+2ab\cos\theta)^{\frac{3}{2}}-
2m(a^2+b^2+2ab\cos\theta)}}\d\theta\wedge \d\varphi.
\end{split}
\eee
Again, denote by $\{\hat{e}_2,\hat{e}_3\}$ the orthonormal frame tangential to the surface $\hat{\Sigma}$. By the chain rule, one has
\bee
\frac{\partial}{\partial\theta}=
-\frac{ab\sin\theta}{\sqrt{a^2+b^2+2ab
\cos\theta}}\frac{\partial}{\partial \hat{R}}
-\frac{b(b+a\cos\theta)}{{a^2+b^2
+2ab\cos\theta}}\frac{\partial}{\partial\Theta}.
\eee
Then
\bee
\begin{split}
\hat{e}_2&=\frac{1}{\sqrt{\hat{\sigma}_{22}}}\bigg{(}
-\frac{ab\sin\theta}{\sqrt{a^2+b^2
+2ab\cos\theta}}\frac{\partial}{\partial \hat{R}}
-\frac{b(b+a\cos\theta)}{{a^2+b^2
+2ab\cos\theta}}\frac{\partial}{\partial\Theta}\bigg{)},\\
\hat{e}_3&=\frac{1}{\sqrt{\hat{\sigma}_{33}}}\frac{\partial}{\partial\varphi}.
\end{split}
\eee

In terms of the orthonormal frame \eqref{sphon},
\bee
\begin{split}
   \hat{e}_2 &=   (\hat{e}_2)^1 \check{e}_1+(\hat{e}_2)^2 \check{e}_2,\\
     \hat{e}_3&=\check{e}_3
\end{split}
\eee
where
\bee
\begin{split}
   (\hat{e}_2)^1 &= -\frac{a\sin\theta\Big(a^2+b^2+2a
b\cos\theta\Big)^{\frac{1}{4}}}{
\bigg{(}(a^2+b^2+2ab\cos\theta)^{\frac{3}{2}}-2m(b+a
\cos\theta)^2\bigg{)}^{\frac{1}{2}}},\\
    (\hat{e}_2)^2&=-\frac{(b+a\cos\theta)\sqrt{(a^2+b^2+2a
b\cos\theta)^{\frac{1}{2}}-2m}}{
\bigg{(}(a^2+b^2+2ab\cos\theta)^{\frac{3}{2}}-2m(b+a
\cos\theta)^2\bigg{)}^{\frac{1}{2}}}.
\end{split}\eee

Then
\be\label{R2323}
\begin{split}
R(\hat{e}_2,\hat{e}_3,\hat{e}_2,\hat{e}_3)
&=R((\hat{e}_{2})^1\check{e}_1
+(\hat{e}_{2})^2\check{e}_2,\check{e}_3,
(\hat{e}_{2})^1\check{e}_1+(\hat{e}_{2})^2
\check{e}_2,\check{e}_3)\\
&=\frac{m}{r^3}\Big(2((\hat{e}_{2})^2)^2-
((\hat{e}_{2})^1)^2\Big)\\
&=\frac{m \left(\sqrt{a^2+2 a b \cos \theta +b^2}-2 m\right) \left(4 (a \cos \theta +b)^2-\frac{2 a^2 \sin ^2\theta }{1-\frac{2 m}{\sqrt{a^2+2 a b \cos \theta +b^2}}}\right)}{2 \left(a^2+2 a b \cos \theta +b^2\right)^{3/2} \left(\left(a^2+2 a b \cos \theta +b^2\right)^{3/2}-2 m (a \cos \theta +b)^2\right)}.\\
&=\frac{m \bigg{(}4(a\cos\theta+b)^2(\sqrt{a^2+2 a b \cos \theta +b^2}-2m)-2a^2\sin^2\theta\sqrt{a^2+2 a b \cos \theta +b^2}\bigg{)}}{2 \left(a^2+2 a b \cos \theta +b^2\right)^{3/2} \left(\left(a^2+2 a b \cos \theta +b^2\right)^{3/2}-2 m (a \cos \theta +b)^2\right)}.
\end{split}
\ee

Finally, the Hayward energy of $\hat{\Sigma}$ can be rewritten as
\be\label{massex}
\begin{split}
\mhy(\hat{\Sigma})&=\frac{1}{8\pi}\sqrt{\frac{|\hat{\Sigma}|}{16\pi}}\int_{\hat{\Sigma}} 2 R(\hat{e}_2,\hat{e}_3,\hat{e}_2,\hat{e}_3)\d\hat{\sigma}\\
&=\frac{1}{4}\sqrt{\frac{|\hat{\Sigma}|}{16\pi}}\int_0^{2\pi}
 \frac{b m (a+b \cos \theta ) \left(4 (a \cos \theta +b)^2-\frac{2 a^2 \sin ^2\theta}{1-\frac{2 m}{\sqrt{a^2+2 a b \cos \theta +b^2}}}\right)}{\left(a^2+2 a b \cos \theta +b^2\right)^2 \sqrt{\frac{\left(a^2+2 a b \cos \theta +b^2\right)^{3/2}-2 m (a \cos \theta +b)^2}{\sqrt{a^2+2 a b \cos \theta +b^2}-2 m}}}\d\theta.\\
\end{split}
\ee

Let us expand the integrand in the energy expression \eqref{massex} for small $b$,
\bee
\frac{b m (a+b \cos \theta ) \left(4 (a \cos \theta +b)^2-\frac{2 a^2 \sin ^2\theta)}{1-\frac{2 m}{\sqrt{a^2+2 a b \cos \theta +b^2}}}\right)}{\left(a^2+2 a b \cos \theta +b^2\right)^2 \sqrt{\frac{\left(a^2+2 a b \cos \theta +b^2\right)^{3/2}-2 m (a \cos \theta +b)^2}{\sqrt{a^2+2 a b \cos \theta +b^2}-2 m}}}  =\Big(\frac{m(4a^2\cos^2\theta-\frac{2a^2\sin^2\theta}{1-\frac{2m}{a}})}{a^3\sqrt{\frac{a^3-2a^2m\cos^2\theta}{a-2m}}}\Big)b+\mathcal{O}(b^2).\eee

Setting $a=2.1m$, numerical integral shows that
\bee
\int_0^{2\pi}\Big(\frac{m(4a^2\cos^2\theta-\frac{2a^2\sin^2\theta}{1-\frac{2m}{a}})}{a^3\sqrt{\frac{a^3-2a^2m\cos^2\theta}{a-2m}}}\Big)\d \theta= -\frac{6.46}{m}<0. \eee

This indicates that in the outer neighborhood of the Schwarzschild horizon ($\hat{R}=2m=4\beta$), `thin tori' have strictly negative Hayward energy. We can indeed to prove the existence of many of such `thin tori' with strictly negative energy in the following analytic way. Let
\be\label{location}
a=(2+2\lambda)m \ \ \mbox{and}\ \ b=\lambda m,
\ee
the curvature term \eqref{R2323} yields that
\bee
\lim_{\lambda\rightarrow 0} R(\hat{e}_2,\hat{e}_3,\hat{e}_2,\hat{e}_3)=
-\frac{1}{8m^2}.\eee
If one takes $\lambda$ sufficiently small, the integrand in the energy expression \eqref{massex} is strictly negative and therefore the Hayward energy of such a torus is negative. According to \eqref{location}, the smallness of the parameter $\lambda$ means both the torus is `very thin' and its location is in the outer neighborhood of the Schwarzschild horizon.

\begin{figure}[htp]
\centering \mbox{{\includegraphics[width=2.0in]{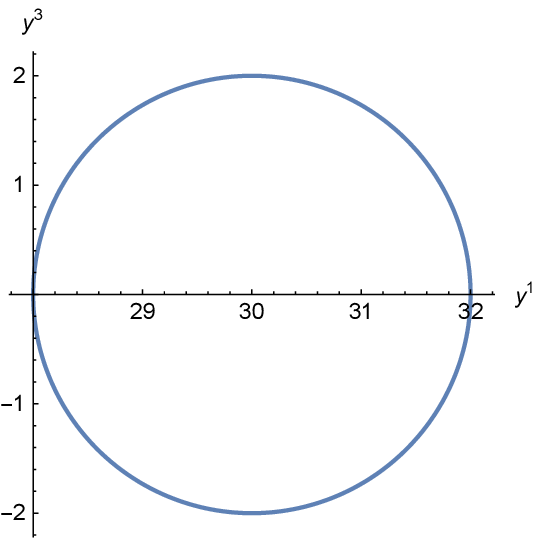}}\quad\ \ \ \ \
{\includegraphics[width=2.2in]{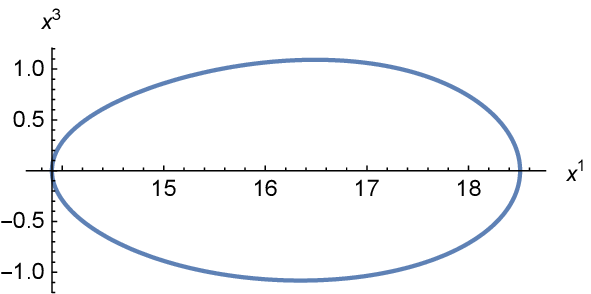}}}\\
\hspace{0mm} (i)  \hspace{58mm} (ii)   \\
\caption{ Graphs of the circle \eqref{circleequation} in different coordinates. For $a=30$, $b=2$ and $m=2\beta=6+4\sqrt{2}$, (i) is the graph of the circle \eqref{circleequation} in the $y^1-y^3$-plane; (ii) is the graph of the circle \eqref{circleequation} in the $x^1-x^3$-plane.  }
\label{transformation}
\end{figure}

The relation of the change of coordinates between the isotropic coordinate and the static coordinate is
\bee
\hat{R}=R(1+\frac{m}{2R})^2
\eee
and further the relation between the Cartesian coordinates $\{x^i \}$ and $\{y^i\}$ is
\bee
\begin{split}
x^1&=\frac{y^1}{2}\bigg(1+\Big(1-\frac{2m}{\hat{R}}
\Big)^{\frac{1}{2}}
-\frac{m}{\hat{R}}\bigg),\\
x^2&=\frac{y^2}{2}\bigg(1+\Big(1-\frac{2m}{\hat{R}}
\Big)^{\frac{1}{2}}
-\frac{m}{\hat{R}}\bigg),\\
x^3&=\frac{y^3}{2}\bigg(1+\Big(1-\frac{2m}{\hat{R}}
\Big)^{\frac{1}{2}}
-\frac{m}{\hat{R}}\bigg).
\end{split}
\eee
If one transforms the standard circle
\be\label{circleequation}
\{(y^1,y^3)|(y^1-a)^2+(y^3)^2=b^2\}
 \ee
 to the $\{x^i\}$ coordinates, the circle is horizontally dragged, as illustrated in Fig. \ref{transformation}.

\begin{figure}[htp]
\begin{center}
\includegraphics[width=0.35\textwidth]{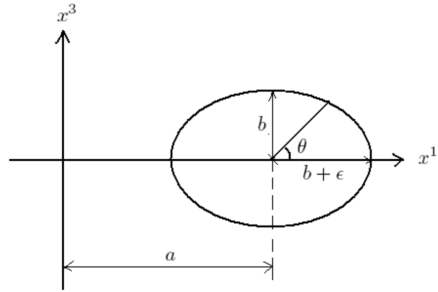}
\caption{\label{horizonaltorus} Horizontally dragged circle in the $x^1-x^3$-plane. }
\end{center}
\end{figure}

This motivates us to conjecture that the horizontal dragging effect may lead to negative contribution to the Hayward energy. The
$\epsilon$-horizontally dragged torus $\bar{\Sigma}$   in the isotropic coordinates $\{x^i\}$  is swept out by rotating an ellipse along the $x^3$-axis, as illustrated in Fig. \ref{horizonaltorus}.

More precisely, the horizonally dragged torus can be parameterized as
\be
\begin{split}
x^1&=(a+(b+\epsilon)\cos\theta)\cos\varphi,\\
x^2&=(a+(b+\epsilon)\cos\theta)\sin\varphi, \ \theta \in [0,2\pi), \ \varphi\in [0,2\pi)\\
x^3&=b\sin\theta.\label{horizonaltoriiso}
\end{split}\ee

It can be obtained from (\ref{horizonaltoriiso}) that
the induced 2-metric $\bar{\sigma}$ on $\bar{\Sigma}$ is
\bee
\bar{\sigma}=\bar{\sigma}_{22}d\theta^2
+\bar{\sigma}_{33}d\varphi^2,
\eee
where
\bee
\begin{split}
\bar{\sigma}_{22}&=\frac{\left(2 b^2-\epsilon  (2 b+\epsilon )\cos 2 \theta +2 b \epsilon +\epsilon ^2\right) \left(\sqrt{2} \sqrt{2 a^2+4 a \cos \theta (b+\epsilon )+2 b^2+\epsilon  (2 b+\epsilon )\cos 2 \theta+2 b \epsilon +\epsilon ^2}+2 \beta\right)^4}{8 \left(2 a^2+4 a \cos \theta (b+\epsilon )+2 b^2+\epsilon  (2 b+\epsilon )\cos 2 \theta+2 b \epsilon +\epsilon ^2\right)^2},\\
\bar{\sigma}_{33}&=\frac{(a+ (b+\epsilon )\cos\theta)^2 \left(\sqrt{2} \sqrt{2 a^2+4 a (b+\epsilon )\cos \theta+2 b^2+\epsilon   (2 b+\epsilon )\cos2\theta+2 b \epsilon +\epsilon ^2}+2\beta\right)^4}{4 \left(2 a^2+4 a (b+\epsilon )\cos \theta+2 b^2+\epsilon   (2 b+\epsilon )\cos 2 \theta+2 b \epsilon +\epsilon ^2\right)^2}.
\end{split}
\eee

Denote by $\{\bar{e}_2,\bar{e}_3\}$ the orthonormal frame tangential to the surface $\bar{\Sigma}$.
By the chain rule, one has
\bee
\begin{split}
\frac{\partial}{\partial\theta}=&
-\frac{(a(b+\epsilon)+\epsilon(2b+\epsilon)\cos\theta)
\sin\theta}{
\sqrt{a^2+b^2+2a(b+\epsilon)\cos\theta+
\epsilon(2b+\epsilon)\cos^2\theta}}\frac{\partial}{\partial R}\\
&-\frac{2b(b+\epsilon+a\cos\theta)}{
2a^2+2b^2+2b\epsilon+\epsilon^2+
4a(b+\epsilon)\cos\theta
+\epsilon(2b+\epsilon)\cos2\theta}
\frac{\partial}{\partial\Theta}.
\end{split}
\eee
Then we have
\bee
\begin{split}
\bar{e}_2=&\frac{1}{\sqrt{\bar{\sigma}_{22}}}\bigg{[}
-\frac{(a(b+\epsilon)+\epsilon(2b+\epsilon)\cos\theta)
\sin\theta}{
\sqrt{a^2+b^2+2a(b+\epsilon)\cos\theta+
\epsilon(2b+\epsilon)\cos^2\theta}}\frac{\partial}{\partial R}\\
&-\frac{2b(b+\epsilon+a\cos\theta)}{
2a^2+2b^2+2b\epsilon+\epsilon^2+
4a(b+\epsilon)\cos\theta
+\epsilon(2b+\epsilon)\cos2\theta}
\frac{\partial}{\partial\Theta}
\bigg{]}\\
&=(\bar{e}_2)^{R}\frac{\partial}{\partial R}+
(\bar{e}_2)^{\Theta}\frac{\partial}{\partial\Theta},\\
\bar{e}_3=&\frac{1}{\sqrt{\bar{\sigma}_{33}}}\frac{
\partial}{\partial\varphi}=(\bar{e}_3)^{\varphi}\frac{\partial}{\partial\varphi}.
\end{split}
\eee
and
\bee
\begin{split}
R(\bar{e}_2,\bar{e}_3,\bar{e}_2,\bar{e}_3)&=\big( (\bar{e}_2)^R \big)^2 \big( (\bar{e}_3)^\varphi \big)^2 R_{R\varphi R\varphi}+ \big( (\bar{e}_2)^\Theta \big)^2 \big( (\bar{e}_3)^\varphi \big)^2 R_{\Theta\varphi \Theta\varphi}\\
&=\frac{16 \sqrt{2} \beta  \sqrt{2 a^2+4 a \cos \theta (b+\epsilon )+2 b^2+\epsilon (2 b+\epsilon ) \cos 2 \theta+2 b \epsilon +\epsilon ^2}}{\left(2 b^2-\epsilon  (2 b+\epsilon )\cos 2 \theta+2 b \epsilon +\epsilon ^2\right) }\cdot\\
&\left(\sqrt{2} \sqrt{2 a^2+4 a \cos \theta (b+\epsilon )+2 b^2+\epsilon  (2 b+\epsilon )\cos 2 \theta+2 b \epsilon +\epsilon ^2}+2 \beta \right)^{-6}\cdot\\
&\Big(4 a^2 \cos 2 \theta  \left(3 b^2+2 b \epsilon +\epsilon ^2\right)+4 a^2 b^2-8 a^2 b \epsilon -4 a^2 \epsilon ^2+4 a \epsilon ^3 \cos 3 \theta +8 a b^2 \epsilon  \cos 3 \theta \\
&+4 a \cos \theta \left(8 b^3+6 b^2 \epsilon -3 b \epsilon ^2-\epsilon ^3\right)+12 a b \epsilon ^2 \cos (3 \theta )+\epsilon ^4 \cos4 \theta\\
&+16 b^4+32 b^3 \epsilon +4 b^2 \epsilon ^2 \cos4 \theta +12 b^2 \epsilon ^2+4 b \epsilon ^3 \cos4 \theta-4 b \epsilon ^3-\epsilon ^4\Big).
\end{split}
\eee

Setting $\beta=3+2\sqrt{2}, a=3.5, b=0.5$ and $\epsilon=1$, numerical result shows that
$\mhy(\bar \Sigma)=-3.17526$.

\begin{figure}[htp]
\begin{center}
\includegraphics[width=0.60\textwidth]{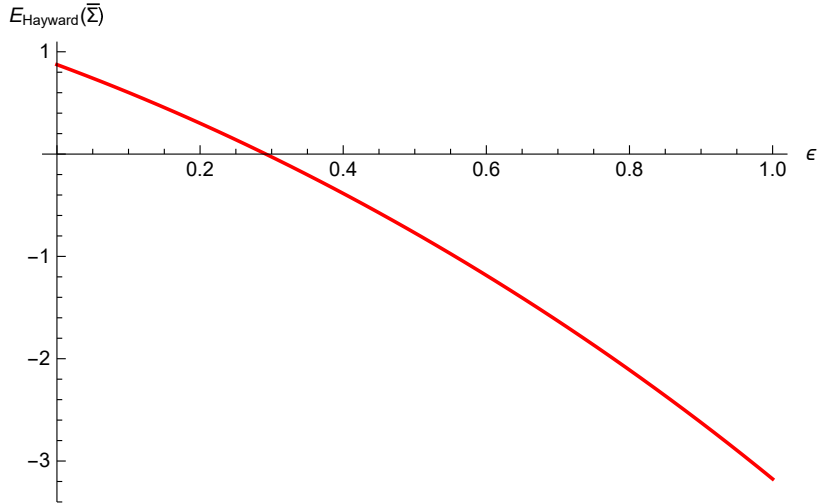}
\caption{\label{isoepsilon} Graph of $\mhy(\bar{\Sigma})$ with respect to $\epsilon$ for  $a=3.5, b=0.5$ and $\beta=3+2\sqrt{2}$. }
\end{center}
\end{figure}

To see the effect of the horizonal dragging parameter $\epsilon$ to the Hayward energy, we plot the value of $\mhy(\bar\Sigma)$ for different $\epsilon$ with fixed $a$, $b$ and $\beta$ in Fig. \ref{isoepsilon}.

\begin{figure}[htp]
\begin{center}
\includegraphics[width=0.50\textwidth]{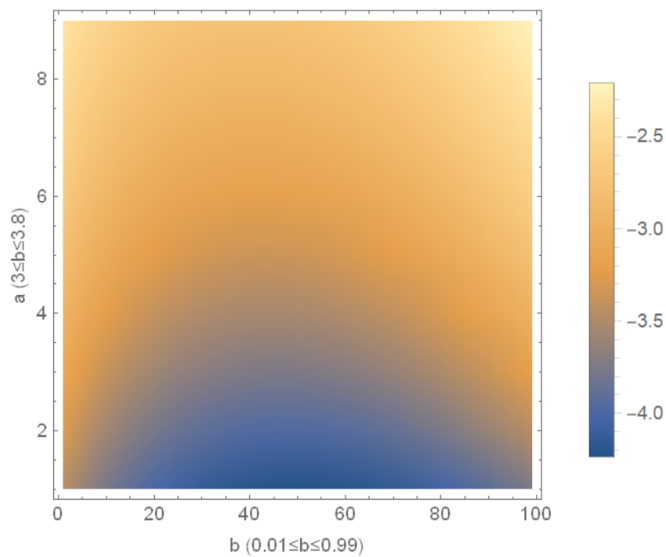}
\caption{\label{isoepsilon=1} Plot of $\mhy(\bar\Sigma)$ with respect to $a$ and $b$ for $\epsilon=1$ and $\beta=3+2\sqrt{2}$. }
\end{center}
\end{figure}

We also plot $\mhy(\bar{\Sigma})$ for different $a$ and $b$ with fixed $\epsilon=1$ in Fig. \ref{isoepsilon=1}.

Then we have numerically found many distorted tori with negative Hayward energy in the region between the boundary of the star and the Schwarzschild horizon. They are horizontally dragged.

\section{Summary}
We consider a scenario of a spherically symmetric constant density star matched to an exterior Schwarzschild solution. The positivity of the Hayward quasilocal energy of tori is investigated. For generic tori entirely within the star, distorted or not, trapped or not, they must have positive Hayward energy. The value of the energy is proportional to the mass density.

Standard `thin tori' in the isotropic coordinates in the exterior Schwarzschild region also have positive Hayward energy. It is nature to conjecture that the Hayward quasilocal energy of tori is always positive in our scenario. However, in the static coordinates, examples of tori with negative Hayward energy have been found in both analytic and numerical ways. They are located in the outer neighborhood of the Schwarzschild horizon. These tori are swept out by the standard circles in the static coordinates but they look horizontally dragged in the isotropic coordinates. The horizontal dragging effect to the negative contribution to the Hayward energy is revealed for the finding of distorted tori with negative Hayward energy in the region between the boundary of the star and the Schwarzschild horizon. The results are not so disastrous. When the major radius of the torus is sufficiently large and the torus goes to spatial infinity, the Hayward energy becomes positive. In fact, for large $a$, one has $\mhy(\Sigma)=\frac{\beta}{2}(\frac{b\pi}{a} )^{\frac{3}{2}} +\mathcal{O}(a^{-\frac{5}{2}})$. The physical significance of these examples of tori with negative energy is that the local dominant energy density, i.e., the scalar curvature of the induced 3-metric on the time symmetric inial data being nonnegative, does not necessarily guarantee the positivity of the quasilocal energy.

The metric is discontinuous at the boundary of the star where the curvatures have a jump. There are marginally trapped tori numerically constructed which are partially inside the star and partially outside the star \cite[Section IV]{KMMOX17}.  They are swept out by a distorted circle and both the induced 2-metric and the unit normal are nonsmooth across the boundary of the star. There should be an influence of the gravitational action at the nonsmooth boundary and certain subtle constraints should be imposed appropriately \cite{HayG93}. It is valuable to seek a framework to defining a notion of quasilocal energy for surfaces with corners. But clearly it is beyond the scope of the current manuscript.

\section*{Acknowledgments}  X. He is partially supported by the Natural Science Foundation of Hunan Province (Grant 2018JJ2073). N. Xie is partially supported by the National Natural Science Foundation of China (Grant 11671089).


\begin{thebibliography}{99}
\bibitem{ADM} R.~Arnowitt, S.~Deser, and C.~Misner, `The dynamics of general relativity', pp.227-265, in \textit {Gravitation: an introduction to current research}, Ed.~L.~Witten, (Wiley, NY, 1962).

\bibitem{Ber} G.~Bergqvist, Enegry of small surfaces, Class. Quantum Grav. {\bf 11}, 3013 (1994).

  \bibitem{BY} J.~D.~Brown and J.~W.~York, Quasilocal energy and conserved charges derived from the gravitational action, Phys. Rev. D {\bf 47}, 1407 (1993).
  \bibitem{ChY88}D.~Christodoulou and S.~-T.~Yau, `Some remarks on the quasi-local mass', pp. 9-14, in \textit{Mathematics and general relativity (Santa Cruz, CA, 1986)}, Ed.~J.~Isenberg, Contemp. Math. {\bf 71}, Amer. Math. Soc., (Providence, RI, 1988).
      \bibitem{FK11} X.~-Q.~Fan and K.~-K.~Kwong, The Brown-York mass of revolution surfaces in asymptotically Schwarzschild manifolds, J. Geom. Anal. {\bf 21}, 527 (2011).
\bibitem{Ge73} R.~Geroch, Energy extraction, Ann. N.Y. Acad. Sci. {\bf 224}, 108 (1973).
    \bibitem{Ha} S.~W.~Hawking, Gravitational radiation in an expanding universe, J. Math. Phys. {\bf 9}, 598 (1968).
    \bibitem{HayG93} G. Hayward, Gravitational action for spacetimes with nonsmooth boundaries, Phys. Rev. D {\bf 47}, 3275 (1993).
\bibitem{SH1} S.~Hayward, Quasilocal gravitational energy, Phys. Rev. D {\bf 49}, 831 (1994).
\bibitem{HWX20} X.~He, L.~F.~Wong, and N.~Xie, On limit behavior of quasi-local mass for ellipsoids at spatial infinity, Commun. Theor. Phys. {\bf 72}, 015403 (2020).
\bibitem{HI01} G.~Huisken and T.~Ilmanen, The inverse mean curvature flow and the Riemannian Penrose inequality, J. Diff. Geom. {\bf 59}, 353 (2001).
\bibitem{Hu96} S.~Husa, Initial data for general relativity containing a marginally outer trapped torus, Phys. Rev. D {\bf 54}, 7311 (1996).
\bibitem{KMMOX17} J.~Karkowsi, P.~Mach, E.~Malec, N.~\'{O} Murchadha, and N.~Xie, Toroidal trapped surfaces and isoperimetric inequalities, Phys. Rev. D {\bf 95}, 064037 (2017).
\bibitem{MX17} P.~Mach and N.~Xie, Toroidal marginally outer trapped surfaces in closed Friedmann-Lema\^{i}tre-Robertson-Walker spacetimes: stability and isoperimetric inequalities, Phys. Rev. D {\bf 96}, 084050 (2017).
 \bibitem{O86} N.~\'{O} Murchadha, How large can a star be?, Phys. Rev. Lett. {\bf 57}, 2466 (1986).
\bibitem{Pe82}R.~Penrose, `Some unsolved problems in classical general relativity', pp. 631-668, in \textit{Seminar on Differential Geometry}, Ed.~S.~-T.~Yau, Ann. Math. Stud. {\bf 102}, (Princeton Univ. Press, NJ, 1982).

    \bibitem{S+YI} R.~Schoen and S.~-T.~Yau, On the proof of the positive mass conjecture in general relativity, Commun. Math. Phys. {\bf 65}, 45 (1979).
\bibitem{S+YII} R.~Schoen and S.~-T.~Yau, Proof of the positive mass theorem. II, Commun. Math. Phys. {\bf 79}, 231 (1981).
\bibitem{ST02} Y.~Shi and L.~-F.~Tam, Positive mass theorem and the boundary behaviors of compact manifolds with nonnegative scalar curvature, J. Diff. Geom. {\bf 62}, 79 (2002).
 \bibitem{Sz} L.~Szabados, Quasi-local energy-momentum and angular momentum in general relativity, Living Rev. Relativity {\bf 12}, 4 (2009).
\bibitem{Wi} E.~Witten, A new proof of the positive energy theorem, Commun. Math. Phys. {\bf 80}, 381 (1981).









 \end{thebibliography}
\end{document}